\documentclass[twocolumn,showpacs,aps,prl,superscriptaddress,floatfix,letter]{revtex4}

\usepackage{epsfig}
\usepackage{graphics}
\usepackage{graphicx}
\usepackage{epic}
\usepackage{floatflt}
\usepackage{dcolumn}
\usepackage{amsmath}

\newcommand{\btodkpi}{\ensuremath{B^0 \to D^\mp K^0 \pi^\pm}}

\newcommand{\de}{\ensuremath{\mbox{$\Delta E$}}}
\newcommand{\fis}{\ensuremath{\mbox{$\mathcal{F}$}}}
\newcommand{\Bbar}{\ensuremath{\overline{B} \rule{0ex}{1ex}^0}}

\newcommand{\tbpg}{\ensuremath{2\beta + \gamma}}
\newcommand{\btoc}{\ensuremath{b \to c}}
\newcommand{\btou}{\ensuremath{b \to u}}
\newcommand{\mkspi}{\ensuremath{m^2(K_S^0 \pi^{\pm})}}

\newcommand{\btoddordds}{\ensuremath{B^0 \to D^{\mp} D_{(s)}^{\pm}}}
\newcommand{\bbbar}{\ensuremath{B\overline{B}}}
\newcommand{\bbbarra}{\ensuremath{B\overline{B}}}
\def\invfb   {\ensuremath{\mbox{\,fb}^{-1}}\xspace}

\newcommand{\ddbarra}{{{\hbox{D}}{\hbox{\kern -.34cm\lower -.26cm\hbox{${\scriptscriptstyle(}{\scriptstyle -}{\scriptscriptstyle )}$}}}}}

\long\def\inst#1{\par\nobreak\kern 4pt\nobreak
    {\it #1}\par\vskip 10pt plus 3pt minus 3pt}

\def\sss{\scriptscriptstyle}
\def\barpd{{\raise.35ex\hbox{${\sss (}$}}--{\raise.35ex\hbox{${\sss )}$}}}
\def\dbarp{\hbox{$D^{0}$\kern-1.3em\raise1.5ex\hbox{\barpd}}}
\def\dbarpnozero{\hbox{$D$\kern-0.85em\raise1.5ex\hbox{\barpd}}}

\newcommand{\BaBarYear}      {07}
\newcommand{\BaBarNumber}    {065}
\newcommand{\SLACPubNumber}  {13050}

\input ./pubboard/babarsym

\begin{document}


 {\pagestyle{empty}
 \begin{flushleft}
 \babar-PUB-\BaBarYear/\BaBarNumber \\
 SLAC-PUB-\SLACPubNumber \\
 \end{flushleft}
 }

\title{Time-dependent Dalitz plot analysis of {\boldmath $B^0 \to D^{\mp} K^0 
\pi^\pm$} decays.}
%
\author{B.~Aubert}
\author{M.~Bona}
\author{Y.~Karyotakis}
\author{J.~P.~Lees}
\author{V.~Poireau}
\author{X.~Prudent}
\author{V.~Tisserand}
\author{A.~Zghiche}
\affiliation{Laboratoire de Physique des Particules, IN2P3/CNRS et Universit\'e de Savoie, F-74941 Annecy-Le-Vieux, France }
\author{J.~Garra~Tico}
\author{E.~Grauges}
\affiliation{Universitat de Barcelona, Facultat de Fisica, Departament ECM, E-08028 Barcelona, Spain }
% ...
\collaboration{The \babar\ Collaboration}
\noaffiliation

\begin{abstract}
\noindent
We present for the first time a measurement of the weak phase $2\beta+\gamma$ obtained
 from a time-dependent Dalitz plot analysis of \btodkpi\ decays.
 Using a sample of
approximately $347\times 10^{6}$ $B \Bbar$ pairs collected by the \babar\
detector at the PEP-II asymmetric-energy storage rings, we obtain 
$  \tbpg = (83 \pm 53 \pm 20)^{\circ} $ and $ (263 \pm 53 \pm 20)^{\circ} $ assuming the ratio $r$ of the 
\btou\ and  \btoc\ decay amplitudes to be 0.3. The 
magnitudes and phases for the resonances associated with the \btoc\   
transitions are also extracted from the fit.  
\end{abstract}

\maketitle

The weak phase 
$\gamma\equiv \arg{\left( -\frac{V_{{\rm ud}}V_{{\rm ub}}^*}{V_{{\rm cd}}V_{{\rm cb}}^*}\right)}$, where $V_{ij}$ are elements of the
Cabibbo-Kobayashi-Maskawa quark-mixing matrix~\cite{CKM}, is the 
least constrained angle of the unitarity triangle~\cite{UTfitCKMfitter}. 
Over the past few years, several methods~\cite{gammaMethods} have been 
employed to measure $\gamma$ directly in charged $B\to D^{(*)}K^{(*)}$ 
decays~\cite{BaBarBellegamma}, where sensitivity to the weak phase arises 
from interference between the $b\to c$ (favored) and $b\to u$ (suppressed) 
transitions. In addition, decays to two-body final states containing charm have been studied, such as 
 $\Bz\to D^{(*)\mp}\pi^{\pm}$ and $\Bz\to D^{\mp}\rho^{\pm}$~\cite{bib:DPi-DRho} which are sensitive to the weak phase $2\beta + \gamma$ due to $\Bz$-$\Bzb$ mixing. The phase $\beta\equiv \arg \left( {-\frac{V_{\rm cd}V_{\rm cb}^*}{V_{\rm td}{V_{\rm tb}^*}}} \right)$ is well measured in neutral $B$ decays to charmonium final states~\cite{bib:babar_sin2b}. 
  The sensitivity of this method is limited by the ratio  
 $r$ between the $b\to u$ and $b\to c$ transitions, which is expected to 
be very small ($\sim 0.02$).  Three-body $B$ decays have been 
suggested~\cite{bib:aps-ap} as a way to avoid this limitation, since $r$ in these decays could
be as large as $0.4$ in some regions of the Dalitz plot. 

 In this paper we report on the first measurement of the weak phase $2\beta +\gamma$ 
obtained from a time-dependent Dalitz plot analysis of the decay 
$\Bz\to D^{\mp}\Kz\pi^{\pm}$~\cite{bib:nousHepPh} (charge conjugation is implied throughout
the paper). In the decay \btodkpi\, the three body final state is reached predominantly  through  intermediate 
$\Bz {\to} \ddbarra^{**0}  K_S^0$ and   $\Bz {\to} D^{-} 
K^{*+}$ decays. In the first case, $\ddbarra^{**0}$ indicates a $D_0^{*}(2400)$ or a  $D_2^{*}(2460)$ 
 state produced through \btou\ and \btoc\ color-suppressed transitions. In the second case,  $K^{*}(892)$, 
$K_0^{*}(1430)$, $K_2^{*}(1430)$ and $K^{*}(1680)$ are produced through 
\btoc\ tree-level transitions. A small contribution from the \btou\ decay $\Bz {\to} D_s^{*+}(2573) \pi^-$ is also expected.

Defining   $\vec{x}$  as the vector of the two invariant masses squared $m^2(\KS \pi^{\pm})$ and $m^2(D^{\pm} \pi^{\mp})$, the amplitude $A$ at each point $\vec{x}$  of the 
Dalitz plot can be parameterized as a coherent sum of two-body decay matrix 
elements according
to the isobar model~\cite{bib:isobar}:
\begin{equation}
A_{c(u)}(\vec{x})e^{i\phi_{c(u)}(\vec{x})}=\sum_j a_je^{i\delta_j} 
BW(\vec{x};M_j,\Gamma_j,s_j) ,
\label{eqres}
\end{equation}
where $c$ ($u$) indicates the \btoc\ (\btou) transition and $\phi$ is the total strong phase.
Each resonance $j$ is parameterized by a magnitude $a_j$, a phase 
$\delta_j$, and a factor $BW(\vec{x};M_j,\Gamma_j,s_j)$   
giving the Lorentz invariant expression for the matrix element of the 
resonance  as a function of the position $\vec{x}$, the spin $s$, the mass 
$M$, and the decay width $\Gamma$.

The time-dependent probability of  
a \Bz or \Bzb initial state to decay to a final state with  a  $D^+$ or $D^-$  can  be expressed as :
\begin{eqnarray}
P(\vec{x},\Delta t, 
\xi,\eta)=\frac{A_{c}(\vec{x})^2+A_{u}(\vec{x})^2}{2}  \times
\frac{e^{-\frac{|\Delta t|}{\tau_B}}}{4 \tau_B}  \times \nonumber \\
 \{1   - \eta \xi  C(\vec{x}) \cos(\Delta m_d \Delta 
t) 
 +   \xi  S_{\eta}(\vec{x})  \sin (\Delta m_d  \Delta t )\}.  
\label{timelikeli}
\end{eqnarray}
Here:
\begin{eqnarray}
S_{\eta}(\vec{x}) &=& \frac{2 \mathrm{Im} (A_{c}(\vec{x})A_{u}(\vec{x}) e^{i( \tbpg 
)+\eta 
i(\phi_{c}(\vec{x})-\phi_{u}(\vec{x}))})}{A_{c}(\vec{x})^2+A_{u}(\vec{x})^2}, \nonumber \\
C(\vec{x}) &=& 
\frac{A_{c}(\vec{x})^2-A_{u}(\vec{x})^2}{A_{c}(\vec{x})^2+A_{u}(\vec{x})^2},
\end{eqnarray}
where  $\Delta t$ is the difference 
           in proper decay times of the reconstructed meson $B_{\rm rec}$ and 
           the flavor-tagging meson $B_{\rm tag}$, $\xi=+1(-1)$ if the  flavor of the
           $B_{\rm rec}$  is a $\Bz (\Bzb)$ and $\eta = +1(-1)$ if the
           final state contains a $D^+(D^-)$.  We use the world averages for 
           the $\Bz$ lifetime $\tau_B$ and the mass-eigenstate difference 
           $\Delta m_d$~\cite{pdg}. 

Because Eq.~\ref{timelikeli} contains the terms  
$BW^j(\vec{x},m,\Gamma,s)$, which vary over 
the Dalitz plot, we can fit the magnitudes $a_j$ and the phases 
$\delta_j$ of Eq.~\ref{eqres} to determine \tbpg\  with only a two-fold ambiguity~\cite{bib:aps-ap}. 
Most of the sensitivity to 
\tbpg\ is expected to come from the interference between \btou\ 
and \btoc\ transitions leading to  $D^{**0} K_S^0$ final states (with 
expected $r \sim 0.4$), and from the interference of the former 
with the $b \to c$ transition of the decay  $B^0 \to D^{-} K^{*+}$.

The analysis is based on  $347\times 10^{6}$ $B
\Bbar$ pairs collected at the $\Upsilon(4S)$ resonance by the \babar\ 
detector  at the PEP-II
storage rings. 
The detector is described in detail elsewhere~\cite{babar-nim}. 
 In order to estimate signal selection 
efficiencies and to study physics backgrounds, a Monte Carlo (MC) 
simulation based on GEANT4~\cite{geant4} is used.

We reconstruct $D^+$ mesons in the decay mode $K^- \pi^+ \pi^+$. 
The tracks from $D^+$ decay are required to originate from a common vertex,
and the kaon is selected using a likelihood based particle identification (PID) algorithm. 
The $D^+$ candidates are required to have a mass within $\pm 12$ \mevcc\
(2$\sigma$) of the nominal $D^+$ mass \cite{pdg}, where $\sigma$ is the
experimental resolution. 
Oppositely charged tracks from a common vertex are recognized as $\KS$
candidates if they have an invariant  mass within $\pm 7$ \mevcc\
(3$\sigma$) of the nominal $\KS$ mass \cite{pdg} and a transverse flight-length significance 4$\sigma$ greater than zero. The $\pi^-$ candidate is a track for which the PID is inconsistent with its being a kaon or an electron.

To form $B^0$ candidates, each $D^{+}$ candidate is combined with
a $\KS$ candidate and a $\pi^-$ candidate requiring that the 
three particles originate from a common vertex. We reject $B^0$ candidates with   \mkspi\  in the window [3.40, 3.95] ${\rm GeV}^2/c^4$   
in order to remove backgrounds with non-zero \CP\ content arising from \btoddordds \ decays.
 Using the beam energy in the $e^+e^-$ center-of-mass (CM) frame, two kinematic variables are constructed:
the beam-energy substituted mass $\mes  \equiv \sqrt{s/4-{p^*_B}^2}$, and the 
difference between the measured $B^0$ candidate 
energy and the beam energy, $\de \equiv E^*_B - \sqrt{s}/2$. Here $ p^*_B$ and $ E^*_B$ are the momentum and the energy of the $B_{\rm rec}$ in the CM frame respectively.
Candidates with \de\ in the range $[-0.1,0.1]$ \gev 
and \mes\ in the range $[5.24,5.29]$ \gevcc are selected.
We require $\left|\cos{\theta_{\rm B}}\right|$, the absolute value of the cosine of the angle between the 
$B_{\rm rec}$ momentum and the beam axis, be less than 0.85, 
and  $\left|\cos{\theta_{\rm T}}\right|$, the absolute value of the cosine     of the angle between the thrust axis 
of the $B_{\rm rec}$ decay products and the thrust axis of the rest of the event (ROE),
 be less than 0.95, both in the CM frame~\cite{thrust}. 

The difference of proper-time  $\deltat$ of the two $B$s in the event is 
calculated from the measured separation $\deltaz$ between the vertices of the $B_{\rm rec}$ and the  $B_{\rm tag}$ 
 along the beam direction~\cite{bib:babar_sin2b}. We accept events with 
calculated $\Delta t$ uncertainty  less than $2.5~$ps and $|\Delta t|<20~$ps.
The average $\deltat$ resolution is approximately $1.1~$ps.
The flavor of the $B_{\rm tag}$ is identified from particles that do not belong to the 
 $B_{\rm rec}$ using a multivariate algorithm~\cite{bib:babar_sin2b}.
The effective efficiency of the tagging algorithm, defined as $Q = 
\Sigma_k\, \epsilon_k(1-2w_k)^2$, is  $(30.1 \pm 0.5)\%$, where   
$\epsilon_k$ and $w_k$ are the efficiency 
and the mistag probability, respectively, for each of the six tagging categories $k$.
Untagged events contribute to the determination of magnitudes and phases of 
the resonances and they are grouped in a separate  
seventh category corresponding to the case $\xi =0$ in 
Eq.~\ref{timelikeli} and containing about 38\% of the events. 

To further suppress the dominant continuum background, which have a more
jet-like shape than $B \Bbar$ events, we use a linear combination \fis\ 
of
five variables: $L_0=\sum_{i} p_i$, $L_2 =\sum_{i} p_i  |\cos 
\theta_i|^2$, the global thrust of the event, 
  $\left|\cos{\theta_{\rm T}}\right|$, and 
             $\left|\cos{\theta_{\rm B}}\right|$.
 Here, $p_i$ is the momentum and $\theta_i$ is the angle, with respect to
the thrust axis of the $B_{\rm rec}$, of the tracks and clusters of the 
ROE in the CM frame. 
 The coefficients of  \fis\  are chosen to maximize the separation 
between the distributions obtained from 
Monte Carlo simulated signal events and 28 \invfb\ of continuum events collected at a CM energy 40 \mev\
below that of the $\Upsilon(4S)$ resonance (off-resonance), whose energy is rescaled to the energy of the beams.
 The correlations among the set of measured values of the variables  (\mes, \de, \fis) are negligible.  
Since both \fis\ and the flavor-tagging utilize the ROE information, the distribution of \fis\ is correlated with the tagging category. To take into account this correlation, we parameterize the \fis\ distribution for each tagging category separately

Approximately 7\% of selected events contain more
than one reconstructed signal candidate, arising primarily from  multiple $D^+$ candidates. We 
select the one having the $D$-candidate mass nearest to the nominal value~\cite{pdg}.
For simulated signal events, the entire selection chain has an efficiency of 
$(9.9 \pm 0.1)$\%, where the error is statistical only.


To separate signal from background  and to determine their yields, we first perform an unbinned extended maximum likelihood fit to the selected on-resonance data sample  in the variables 
 \mes, \de, and \fis. The role of this 
first step fit is to extract all the shape parameters, the fractions of 
events by tagging category, and the overall yields, which will 
then be fixed in the subsequent time-dependent fit to the Dalitz plot. 
  We define the logarithm of the likelihood:
\begin{eqnarray}
  \ln \mathcal{Y} = \!\!\!
      \sum_{\mbox{\tiny $k$=1}}^{\mbox{\tiny 7}} \left( \sum_{\mbox{\tiny $i$=1}}^{\mbox{\tiny $N_{tot}$}} \ln
      \left( \sum_{\mbox{\tiny $j$}} N_{jk} {Y}_{jk}^{i} \right) -
    \sum_{\mbox{\tiny $j$}} N_{jk}\right),
\label{eq:llh}
\end{eqnarray}
where ${Y}_{jk}^{i}$ is the product of the PDFs of \mes, \de, and \fis$_k$\ for the event $i$ in the tagging category $k$. $N_{tot}$ is the total number of events and $N_{jk}$ is the number of events of
each sample component $j$: signal (Sig), continuum (Cont), 
combinatoric $B \Bbar$ decays ($\bbbar$) and $B \Bbar$ events 
that peak in \mes\ but not in the \de\ signal region (Peak).

The signal is described by a Gaussian function for the \mes distribution, 
two Gaussian functions with common mean for the \de\ distribution, and a 
Gaussian function with different widths on each side of the mean 
(``bifurcated Gaussian
function'') for the \fis\ distribution. The signal model parameters are obtained from a fit to a high-statistics data control sample of   
$B^0 \to D^{\mp} a_1^\pm$ decays. The selection of these events is similar to signal, except that no $\KS$ candidate is 
required. The decay chain  $ a_1^{\pm} \to \rho^0 \pi^{\pm}$ with  
$ 
\rho^0 \to \pi^{\pm} \pi^{\mp} $ is reconstructed requiring the dipion  
invariant mass be within $\pm$ 150 \mevcc of the nominal $\rho^0$ mass~\cite{pdg}, and the 
invariant mass of the $\rho$ candidate with the third pion be within $\pm$ 250 
\mevcc of the nominal $ a_1^{\pm}$ mass~\cite{pdg}.\\
\indent
The \mes\ distributions of the continuum and combinatoric $B \Bbar$ 
backgrounds are described
by empirical threshold functions \cite{argus}, while for \de\ distributions 
linear functions are used.  The \fis\ distributions are parameterized by a  
bifurcated Gaussian function for the continuum background  and a sum of two Gaussian functions for 
the  $B \Bbar$  combinatoric background. For the latter the parameters are determined by  $B \Bbar$ 
Monte Carlo simulation. All the shape parameters of the continuum background are 
taken from fitting the  off-resonance data.
The \mes distribution of the $\mathrm{Peak}$ background is parameterized 
by a Gaussian function with the same mean as the signal and a width fixed 
to the value obtained from Monte Carlo simulation. The \de\ distribution is 
described by an exponential function. The \fis\ 
distribution of $\mathrm{Peak}$ is  described using  
the same PDF as for  $B \Bbar$ background.

The yields and the fraction of events for each tagging category are fitted 
together with the free shape parameters.  The yields obtained for each 
component are 
$N_{\mathrm{Sig}} = 558 \pm 34$, $N_{\mathrm{Cont}} = 13222 \pm 226$, $N_{\bbbar} = 5647 \pm 
213$ and $N_{\mathrm{Peak}} = 183 \pm 41$, in agreement
with the previous result~\cite{bib:babarmh}. 

The second stage of the analysis is the time-dependent fit to the Dalitz plot. 
For each background component, the $\Delta t$ distribution is modeled as an exponential, with an effective lifetime parameter. To model the detector resolution, it is convolved  with the sum of three Gaussians, the sum of two Gaussians and one Gaussian in the case of continuum background, \bbbarra\  combinatorial background and  $\mathrm{Peak}$ respectively.   
The widths of the Gaussians, the relative fraction of them, the effective dilution parameters, and the effective lifetimes are determined independently from fits to the control samples: the off-resonance data sample for the  continuum background, the \bbbarra\  Monte Carlo sample for \bbbarra\ combinatorial background and the $\mathrm{Peak}$  component in the \bbbarra\ Monte Carlo sample for the $\mathrm{Peak}$. In the case of $\mathrm{Peak}$ the 
lifetime is fixed to the  $B^0$ lifetime~\cite{pdg}.
The $\Delta t$  parameterizations described above for each background component are combined in a global time-dependent PDF  ${\mathcal T}^i_{\pm,\rm{Bkgd}}$ obtained  as a weighted average based on  the fitted yields, where $+$ ($-$) indicates $B_{\mathrm{tag}} = B^0$ ($B_{\mathrm{tag}} = \Bzb$).

To obtain the PDF describing the Dalitz plot of the background in the tagging category $k$, we use the results of the yields fit and calculate for each event a background weight \cite{bib:splot}:
\begin{eqnarray}
  W_{\mbox{\tiny Bkgd}}^k = 1 - W_{\mbox{\tiny Sig}}^k \equiv   1 -
    \frac{\sum_j {\mathbf V}_{\mbox{\tiny Sig},j} \: { 
Y}_{jk}(\mes,\de,\fis_k)}
         {\sum_j N_j \: { Y}_{jk}(\mes,\de,\fis_k)},
\end{eqnarray}
where $N_j$ and ${ Y}_{jk}$ are defined as in Eq.~\ref{eq:llh}, and
$\mathbf{V}_{\mbox{\tiny Sig},j}$ is the signal row of the covariance
matrix of the component yields obtained from the likelihood fit.
In the absence of correlations, $W_{\mbox{\tiny Bkgd}}$ are the background 
probabilities $P_{\mbox{\tiny Bkgd}}/P_{\mbox{\tiny total}}$. 
Applying these weights to the Dalitz plot of on-resonance data we obtain the 
observed background Dalitz plot $ {\mathcal D}_{\rm{Bkgd}}$.

For the signal the effect of finite \deltat  
resolution is described by 
convolving Eq.~\ref{timelikeli} with a resolution function composed of 
three Gaussian distributions. 
Incorrect tagging dilutes  the coefficient of $\cos(\deltamd\deltat)$ in 
Eq.~\ref{timelikeli} by a factor 
$(1-2w_i)$. The parameters of the resolution function and those associated with flavor tagging are fixed to the values obtained in  ~\cite{bib:babar_sin2b}.

The expression for the time-dependent Dalitz plot likelihood function is then: 
\begin{eqnarray}
\ln{ {\mathcal L}} =
 \sum_{\mbox{\tiny k=1}}^{\mbox{\tiny 7}}
 \left[
 \sum_{\Bz\ {\rm tag} }
 { \ln{ {\mathcal L}_{+,k} } }
  +
 \sum_{\Bzb\ {\rm tag} }
 { \ln{ {\mathcal L}_{-,k} } }
 \right],
\label{eq:cp_ll_func}
\end{eqnarray}
The likelihood function ${\mathcal 
L}_{+,k}$ (${
\mathcal L}_{-,k}$) for an event in the tagging 
category $k$ with $B_{\mathrm{tag}} = B^0$ ($B_{\mathrm{tag}} = \Bzb$) is:
\begin{eqnarray}
{\mathcal L}_{\pm,k} =  N^k_{\mathrm{Sig}} {\mathcal P}^k_{\pm,\mathrm{Sig}} { 
Y}^k_{\mathrm{Sig}} 
+ N^k_{\mathrm{Bkgd}}  {\mathcal D}_{\mathrm{Bkgd}} {\mathcal T}^k_{\pm,\mathrm{Bkgd}} { 
Y}^k_{\mathrm{Bkgd}}
\label{eq:likelihoodAll}
\end{eqnarray}
Here ${Y}$ indicates the product of PDFs for \mes, \de, and \fis$_k$, 
${\mathcal P}_{\pm,\mathrm{Sig}}$ is the time-dependent Dalitz plot PDF for 
signal. The ${Y}_{Bkgd}$ 
parameterization  is obtained from a weighted average, using the fitted 
yields, of the shapes obtained from the first step fit.

With the current dataset we are
            unable to determine the magnitudes for the suppressed \btou\ decays. We therefore fix the ratio $r= \frac{A(b \to u)}{A(b \to c)}= 0.3$ for each
resonance in the PDF, which is compatible with the limit $r < 0.4$ 
($90\%$ C.L.) reported in Ref.~\cite{bib:babarshahram}.  
The $D_s^{*+}(2573)$ magnitude and phase are fixed to the values given 
in \cite{bib:nousHepPh}. Despite the fact that the \btou\ phases cannot 
be 
precisely determined they are left free in the fit. 
 All the \btoc\ magnitudes 
and phases together with \tbpg\ are free parameters.
 The whole fitting procedure has been validated using high statistic parameterized (toy) Monte Carlo samples. 

\begin{figure}[htpb]
\begin{center}
\setlength{\unitlength}{1mm}
\begin{picture}(80,55)(0,0)
  \jput(-8,0){\epsfig{file=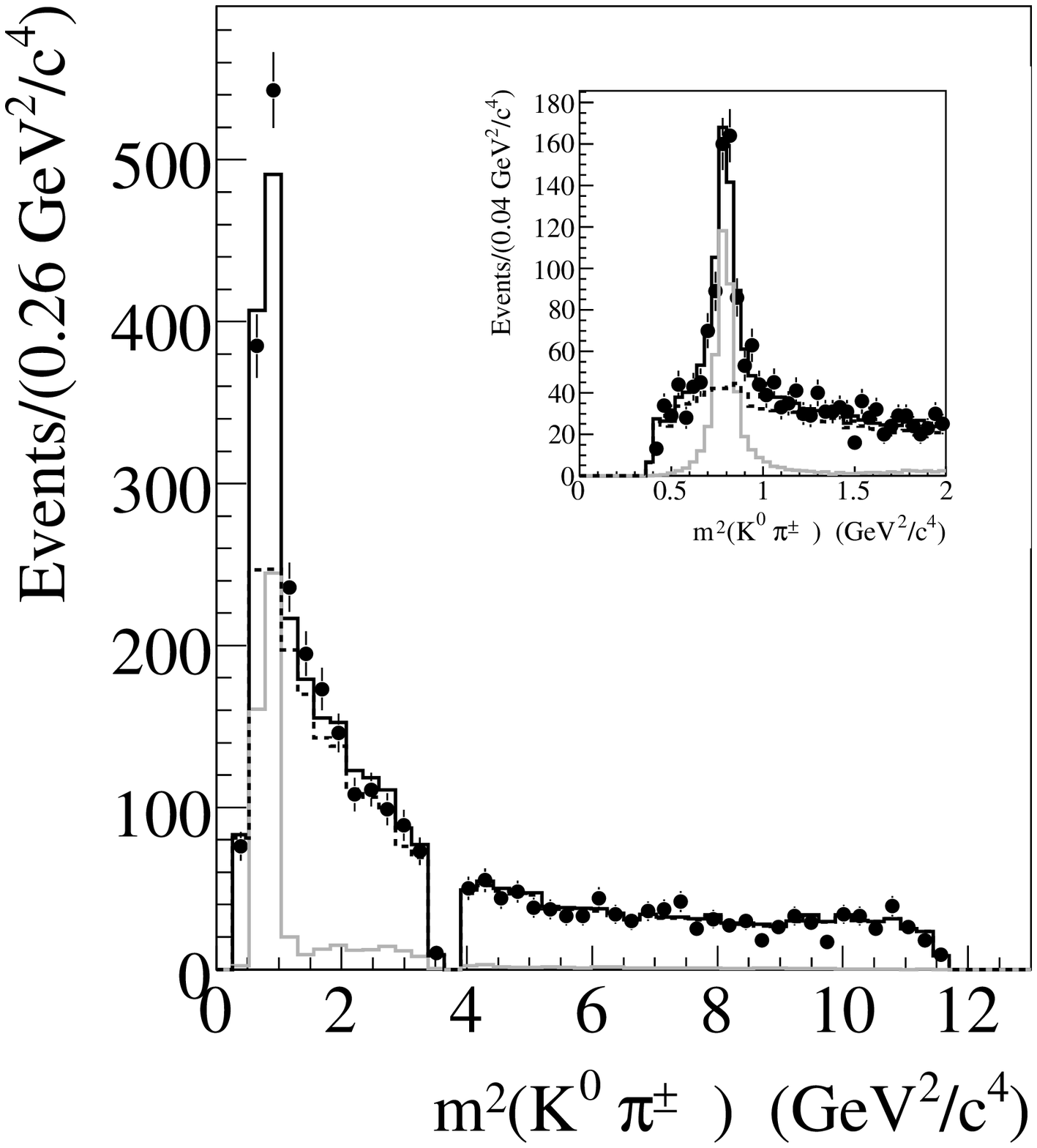,width=52mm}} 
  \jput(41,0){\epsfig{file=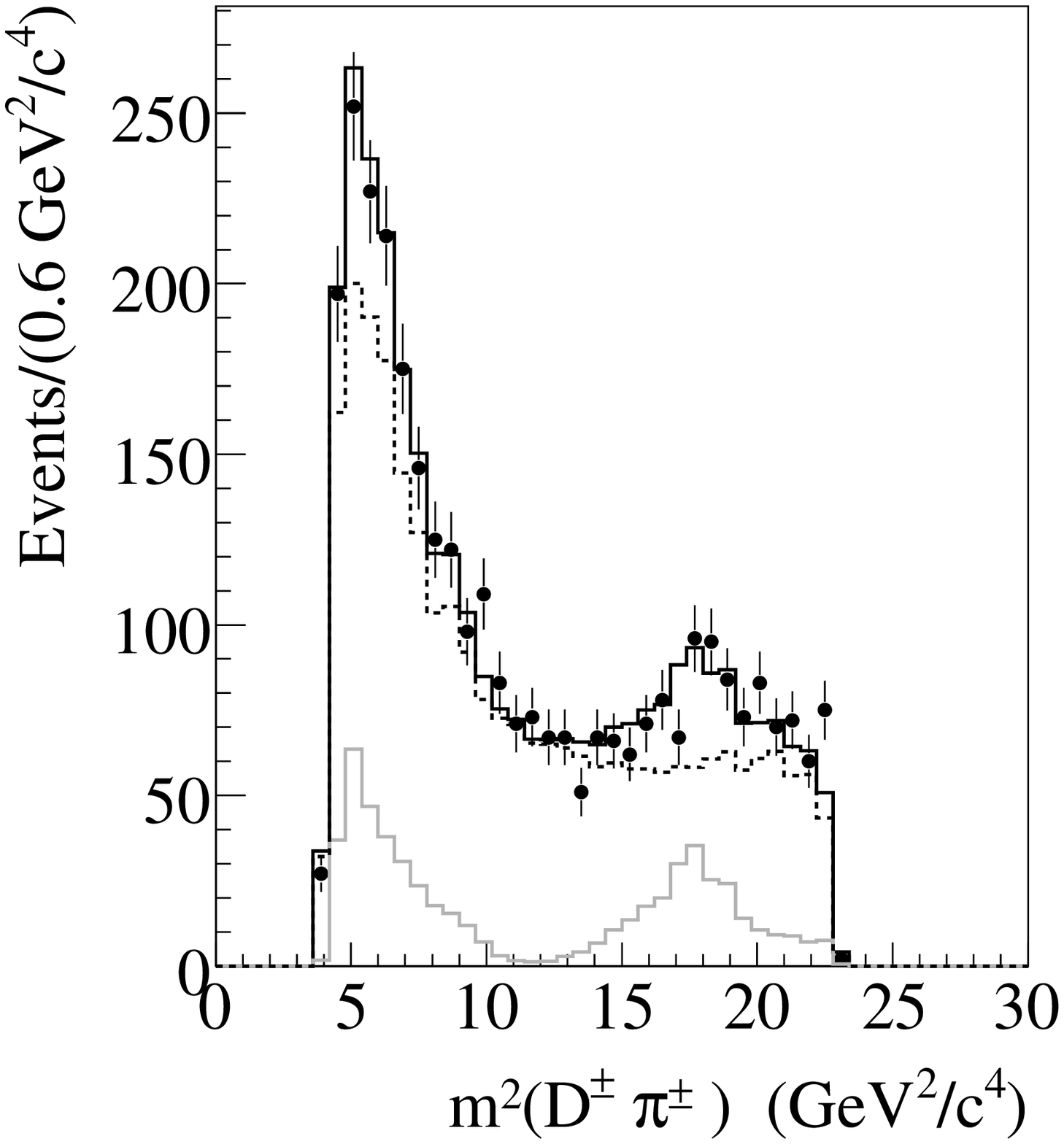,width=52mm}} 
\end{picture}
\caption{The distributions of $m^2(\KS \pi)$ and $m^2(D \pi)$ in data 
(solid points). The overall PDF is superimposed. 
The grey full line is the signal PDF, the dashed line is the background 
PDF.}
\label{fig:Dalitz}
\end{center}
\end{figure}

\begin{figure}[htpb]
\begin{center}
\setlength{\unitlength}{1mm}
\begin{picture}(80,55)(0,0)

\jput(-8,0){\epsfig{file=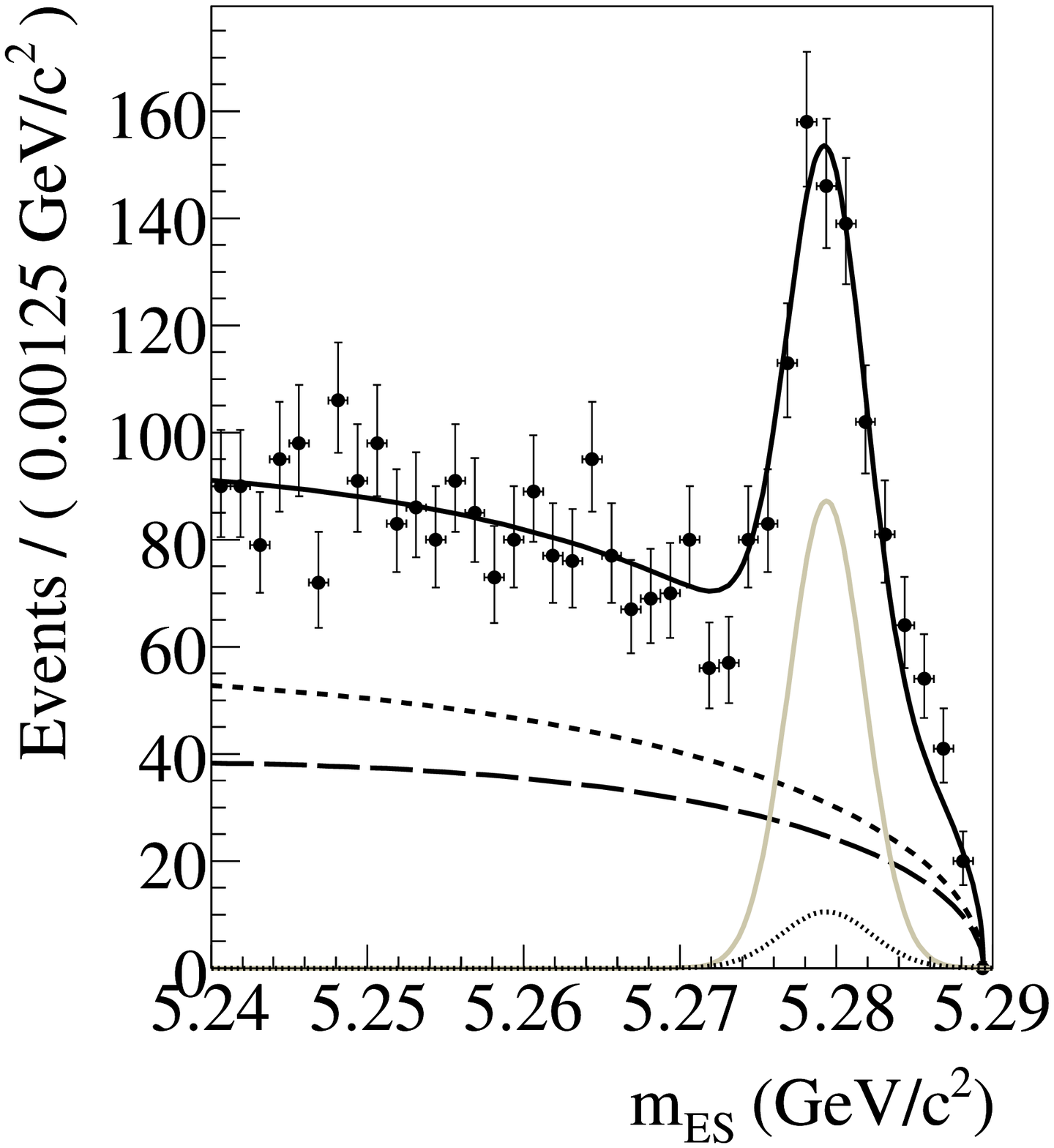,height=52mm,width=50mm}}
  \jput(5,45){a)}
 \jput(39,0){\epsfig{file=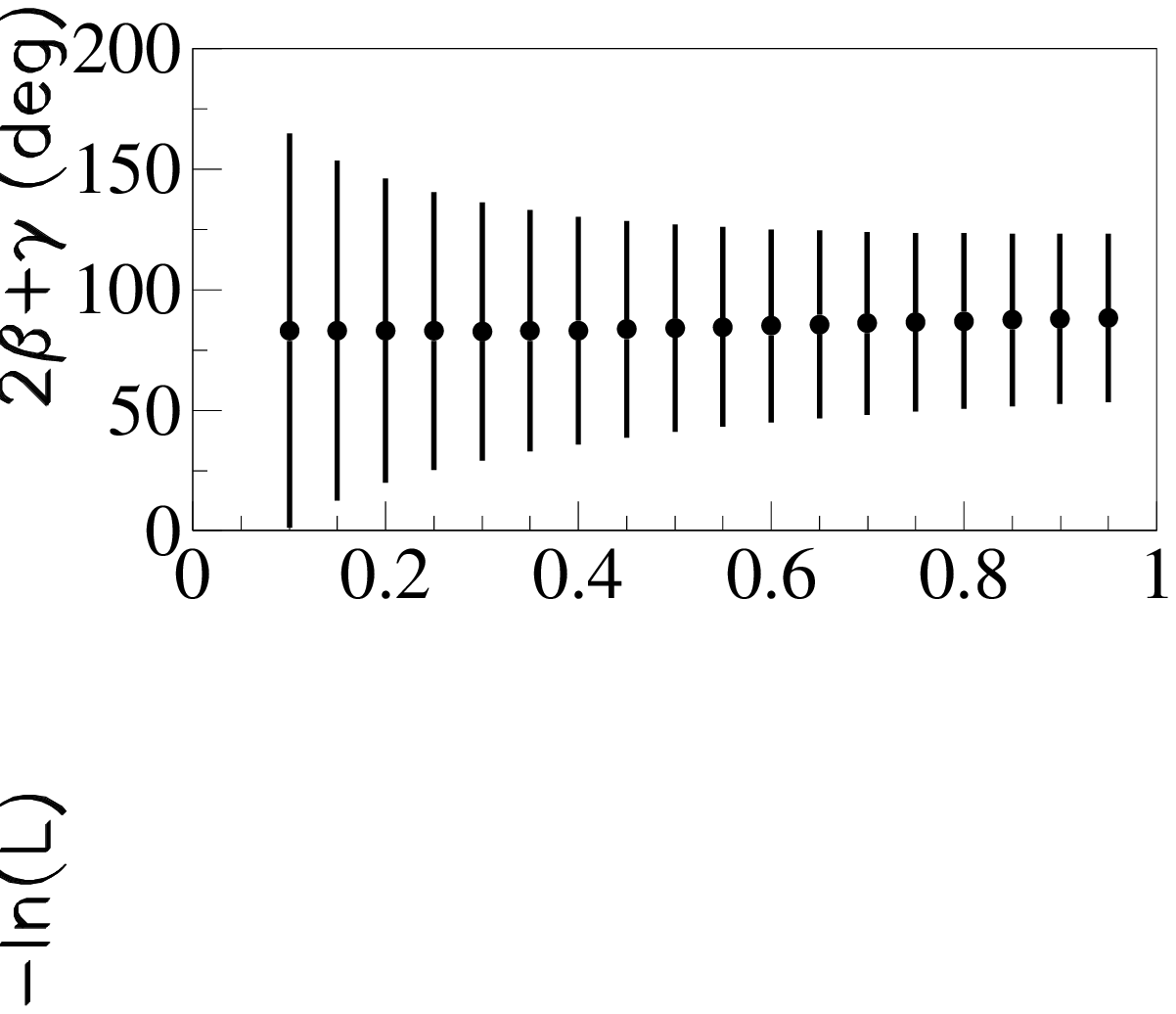,height=56mm,width=50mm}}
  \jput(50,46){b)}
 \jput(40,5){\epsfig{file=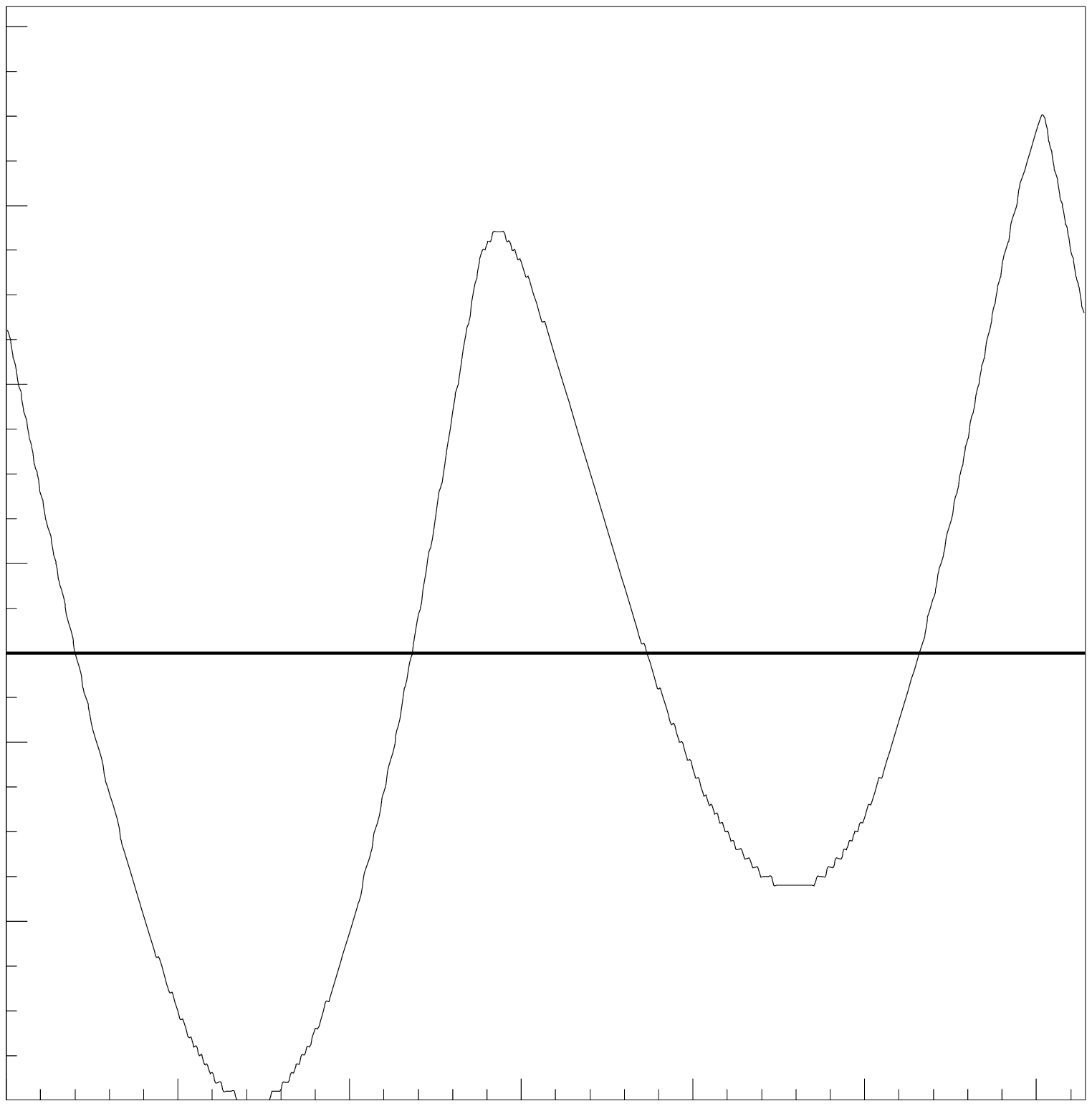,height=22mm,width=47mm}}
  \jput(50,22){c)}
  \scriptsize
  \jput(46,5){0}
  \jput(51.5,5){1}
  \jput(57.,5){2}
  \jput(62.8,5){3}
  \jput(68.5,5){4}
  \jput(74.2,5){5}
  \jput(80.,5){6}
  \jput(65,2){$2 \beta + \gamma$ (rad)}
  \jput(42,7){0}
  \jput(42,10){0.2}
  \jput(42,13){0.4}
  \jput(42,15.8){0.6}
  \jput(42,18.6){0.8}
  \jput(42,21.4){1.0}
  \jput(42,24.2){1.2}
  \jput(80,27){$r$ }
  \normalsize

\end{picture}
\caption{ a): the \mes distribution of on-resonance data (solid points) for the global fitted PDF (blue) with the contribution of the PDF for each component superimposed: signal (grey line), continuum (small dashed line), combinatoric $B \Bbar$ (big dashed line) and Peak (dotted line). To enhance the signal $|\de|<0.025$ \gev and $\fis>0.2$ have been required.
b): distribution of the values of \tbpg\ fitted on
data for different
hypothesis on the $r$ value. c): variation of the logarithm of the
likelihood with \tbpg.}
\label{fig:scanR}
\end{center}
\end{figure}

The fit is performed on events satisfying \mes 
$>5.27$ \gevcc , $|\de|<50$ \mev and $\fis >-2$.   
Results are shown in Table \ref{tab:results}.   Figure \ref{fig:Dalitz} shows the projections of the on-resonance data sample 
on the two Dalitz plot variables $m^2(\KS \pi^{\pm})$ and $m^2(D^{\pm} 
\pi^{\mp})$ with the fitted PDFs superimposed. Figure~\ref{fig:scanR}a shows  the \mes\ distribution and the fitted PDFs for each component, after applying additional requirements on \de\ and \fis. 
 Besides the value of \tbpg, an important outcome of the analysis is the 
fit of the resonance contributions to the \btoc\ part of the Dalitz 
plot. Biases related to the small sample size are observed in the measurement of the magnitudes. They are estimated using a large number of toy experiments generated with the magnitudes values obtained in the fit to the on-resonance data sample.

%
\indent
The systematic errors are summarized in Table \ref{tab:systematics}.
The main contribution is related to the parameterization of the background 
Dalitz plot. This effect has been estimated by 
repeating the fit with a parameterization obtained from off-resonance data 
and $B \bar B$ generic Monte Carlo simulation.  
The systematic uncertainty due to the efficiency variation over the Dalitz 
plot has been evaluated assuming a flat efficiency.  
The effect of potential \CP\ content of the $B \bar B$ peaking background is 
taken into account assuming the same \CP\ violation structure 
as in the signal with a value $r_{\rm{eff}}=0.4$. 
The systematic uncertainties on  the signal Dalitz  plot come from the 
variation of the $r$ factor $(0.3 \pm 0.1)$,  
 of the $D_s^{*+}(2573)$ magnitude  $(0.02 \pm 0.01)$ and from the introduction of up to  7\% of a non resonant component. 
In addition, the masses and widths of the resonances have been varied by 
one standard deviation~\cite{pdg}. We obtain the systematic uncertainty
           arising from imperfect knowledge of the ${\cal Y}$ shape parameters
           and the yields by varying all fixed parameters within their 
           uncertainties.
 Similar variations are applied to the signal 
and 
background fractions in each tagging category as well as for the \deltat 
resolution parameters, the effective lifetimes, the  
$B$ lifetime and the mixing frequency. 
 The systematic uncertainties due to the  
dependence of the tagging efficiency on the $B$ flavor, the 
 beam spot position and the SVT alignment have been obtained following the 
procedure described 
in~\cite{bib:babar_sin2b}. 

Figure~\ref{fig:scanR}b shows the dependence of the measurement of \tbpg\  
on $r$. For each fixed value of $r$, a point in the plot represents the 
result of the fit on 
\tbpg\ with its statistical error. The error decreases, as expected, for 
increasing $r$ and 
the central value remains stable. 
The projection on \tbpg\ of the negative logarithm of the likelihood in Figure~\ref{fig:scanR}c clearly shows the minimum corresponding to the result of the fit and the expected mirror solution at +$\pi$ rad. Having fixed some magnitudes and strong phases, the second solution is disfavored, but it should be regarded as equivalent.

In summary, we present the first results of a time-dependent Dalitz plot analysis 
            of the decay $\Bz\to D^{\pm}\Kz\pi^{\mp}$ to determine the
            Dalitz plot model parameters and the weak phase $2\beta+\gamma$. 
            Assuming $r = 0.3$ we find 
            $2\beta + \gamma = (83\pm 53\pm 20)^{\circ}$  and $(263 \pm 53 \pm 20)^{\circ} $, where the first
            error is statistical and the second is systematic.

\begin{table*}[htbp]
\begin{center}
\leavevmode
\caption{Results for the \btoc\ transitions magnitudes and phases and for \tbpg\ assuming $r=0.3$. The first uncertainty is statistical, the second is 
systematic.} 
\begin{tabular}{l|c|c|c}
\hline
\hline
  Resonance                   &Bias correction          & $V_{cb}$ magnitude after        
& Phase $(^{\circ})$  \\
                              &for the magnitude        & bias correction         
&   \\
\hline
 $K^{*}(892)      $           & $\--$                    &1.                       
& 0.    \\
 $D^{*}_{0}(2400)$           & $+0.003$                &$0.290 \pm 0.048 
\pm 0.067$   & $267 \pm 22 \pm 35$ \\ 
 $D^{*}_{2}(2460)$           & $-0.033$                &$0.042 \pm 0.050 
\pm 0.048$   & $325 \pm 46 \pm 20$ \\ 
 $K^{*}_{0}(1430)$           & $-0.025$                &$0.135 \pm 0.058 
\pm 0.099$   & $284 \pm 30 \pm 11$ \\ 
 $K^{*}_{2}(1430)$           & $-0.017$                &$0.108 \pm 0.056 
\pm 0.051$   & $221 \pm 30 \pm 14$ \\ 
 $K^{*}(1680)    $           & $-0.011$                &$0.404 \pm 0.047 
\pm 0.046$   & $128 \pm 22 \pm 24$ \\ 
\hline
  \tbpg                       & \multicolumn{3}{c}{$(83 \pm 53  \pm 
20)^{\circ}$  and $(263 \pm 53 \pm 20)^{\circ} $}  \\
\hline
\hline
\end{tabular}
\label{tab:results}
\end{center}
\end{table*}

\begin{table*}[htbp]
\begin{center}
\leavevmode
\caption{Sources and sizes of systematic errors.} 
\begin{tabular}{l|c|cc|cc|cc|cc|cc}
\hline
\hline
                        &                       & 
\multicolumn{2}{c|}{$D^{*}_{0}(2400)$}  
& \multicolumn{2}{c|}{$D^{*}_{2}(2460)$}       & 
\multicolumn{2}{c|}{$K^{*}_{0}(1430)$}  
& \multicolumn{2}{c|}{$K^{*}_{2}(1430)$}        & 
\multicolumn{2}{c}{$K^{*}(1680)$} \\  
  Systematic                   & \tbpg\         & $a_c$   &$\delta_c$               
& $a_c$           &$\delta_c$      
& $a_c$   &$\delta_c$              & $a_c$           &$\delta_c$        & 
$a_c$   &$\delta_c$                \\
\hline
Bkgd Dalitz plot param.     &          16.0$^{\circ}$ & 0.058    & 3.2$^{\circ}$
&   0.034               &       12.1$^{\circ}$              &         0.088               
& 9.5$^{\circ}$
&   0.005               &       12.0$^{\circ}$              &         0.015               
& 10.3$^{\circ}$  \\
Eff. over the Dalitz plot      &  5.8$^{\circ}$  & 0.014    & 17.5$^{\circ}$
&    0.028              &         10.8$^{\circ}$            &       0.005                 
& 1.9$^{\circ}$
&    0.036              &         0.8$^{\circ}$             &       0.017                 
& 19.4$^{\circ}$  \\
\CP\ content of bkgd      &   1.0$^{\circ}$   & 0.021    & 6.9$^{\circ}$
&    0.003             &        8.4$^{\circ}$              &         0.005               
& 1.4$^{\circ}$
&    0.007              &       3.9$^{\circ}$               &        0.003                
&  1.0$^{\circ}$ \\
$r$                     &    1.0$^{\circ}$  &  0.013   & 8.6$^{\circ}$
& 0.013                 &      2.2$^{\circ}$                &         0.039               
&  3.0$^{\circ}$
& 0.012                 &      0.7$^{\circ}$                &         0.016               
& 0.3$^{\circ}$  \\
$a(D_s^{*+}(2573))$    &    0.7$^{\circ}$  &  -       & - 
&     -                 &        -                &          -                  
& -
&     -                 &       -                 &          -                  
& -  \\
$m, \Gamma$     &           9.5$^{\circ}$   &  0.012   & 28.0$^{\circ}$
&    0.011              &     6.9$^{\circ}$                 &           0.018             
& 2.8$^{\circ}$
&    0.032              &     5.9$^{\circ}$                 &           0.036             
& 9.3$^{\circ}$  \\
${\mathcal Y}$ PDF param.&  3.0$^{\circ}$  &   0.005   &  1.4$^{\circ}$
&  0.002                &   0.4$^{\circ}$                   &     0.007                   
&   0.6$^{\circ}$
& 0.003                 &   0.1$^{\circ}$                   &    0.002                    
&   0.5$^{\circ}$\\
Signal and bkgd frac. &      1.4$^{\circ}$       &   0.012   &  2.9$^{\circ}$
&  0.004                &    1.2$^{\circ}$                  &       0.013                 
&  1.4$^{\circ}$
&  0.008                &    0.7$^{\circ}$                  &     0.004                   
&  1.4$^{\circ}$\\
Yields  &            0.1$^{\circ}$   &   0.003   &  1.3$^{\circ}$
&  0.001               &    0.3$^{\circ}$                  &      0.005                  
&  0.4$^{\circ}$
&  0.002               &    0.1$^{\circ}$                  &      0.002                  
&  0.1$^{\circ}$ \\
Tagging and time param. &          2.6$^{\circ}$    &   0.003   &  1.4$^{\circ}$
&  0.001                &   0.3$^{\circ}$                   &        0.004                
&  0.4$^{\circ}$
&   0.002               &   0.2$^{\circ}$                  &        0.002               
&  0.2$^{\circ}$ \\
\hline
\hline
\end{tabular}
\label{tab:systematics}
\end{center}
\end{table*}

%


We are grateful for the 
extraordinary contributions of our \pep2\ colleagues in
achieving the excellent luminosity and machine conditions
that have made this work possible.
The success of this project also relies critically on the 
expertise and dedication of the computing organizations that 
support \babar.
The collaborating institutions wish to thank 
SLAC for its support and the kind hospitality extended to them. 
This work is supported by the
US Department of Energy
and National Science Foundation, the
Natural Sciences and Engineering Research Council (Canada),
the Commissariat \`a l'Energie Atomique and
Institut National de Physique Nucl\'eaire et de Physique des Particules
(France), the
Bundesministerium f\"ur Bildung und Forschung and
Deutsche Forschungsgemeinschaft
(Germany), the
Istituto Nazionale di Fisica Nucleare (Italy),
the Foundation for Fundamental Research on Matter (The Netherlands),
the Research Council of Norway, the
Ministry of Science and Technology of the Russian Federation, 
Ministerio de Educaci\'on y Ciencia (Spain), and the
Science and Technology Facilities Council (United Kingdom).
Individuals have received support from 
the Marie-Curie IEF program (European Union) and
the A. P. Sloan Foundation.

---------------------------------------------------------------------------
%
---------------------------------------------------------------------------


\begin{thebibliography}{99}

\bibitem{CKM}
N.Cabibbo, Phys. Rev. Lett. {\bf 10}, 531 (1963); M.Kobayashi and T. Maskawa, Prog. Theor. Phys. {\bf 49}, 652 (1973)

\bibitem{UTfitCKMfitter}
  UTfit Collaboration, M.~Bona {\it et al.}  , 
  JHEP {\bf 0610}, 081 (2006); 
  CKMfitter Group, Charles et al., Eur. Phys. J C41, 1-131.
  arXiv:hep-ph/0410173.

\bibitem{gammaMethods}
                          M. Gronau and D. London,  Phys. Lett. {\bf B253}, 483 (1991);~
                          M. Gronau and D. Wyler,  Phys. Lett.{\bf B265}, 172 (1991);~
                          I. Dunietz,  Phys. Lett. {\bf B270}, 75 (1991);~
                          I. Dunietz,  Z. Phys. {\bf C56}, 129 (1992);~
                          D. Atwood, G. Eilam, M. Gronau and A. Soni,  Phys. Lett. {\bf B341}, 372 (1995);~
                          D. Atwood, I. Dunietz and A. Soni,  Phys. Rev. Lett. {\bf 78}, 3257 (1997);~
                          A. Giri, Yu. Grossman, A. Soffer and J. Zupan, Phys. Rev. {\bf D68}, 054018 (2003).

\bibitem{BaBarBellegamma}
\babar\ Collaboration, B.\ Aubert {\em et al.}, \jprl{95}, 121802 (2005); 
Belle Collaboration, A.~Poluetkov {\em et al.}, \jprd{73}, 112009 (2006).
\babar\ Collaboration, B.\ Aubert {\em et al.}, \jprd{73}, 051105 (2006);
 Belle Collaboration, K. \ Abe {\em et al.}, \jprd{73}, 051106 (2006);
\babar\ Collaboration, B.\ Aubert {\em et al.}, \jprd{71}, 031102 (2005);
\babar\ Collaboration, B.\ Aubert {\em et al.}, \jprd{72}, 071103 (2005);
\babar\ Collaboration, B.\ Aubert {\em et al.}, arXiv:0708:0182 [hep-ex];
\babar\ Collaboration, B.\ Aubert {\em et al.}, \jprd{72}, 032004 (2005);
 Belle Collaboration, M. \ Saigo {\em et al.}, \jprl{94}, 091601 (2005);
\babar\ Collaboration, B.\ Aubert {\em et al.}, \jprd{72}, 071104 (2005);







  \bibitem{bib:DPi-DRho}
  \babar\ Collaboration, B.\ Aubert {\em et al.}, 
Phys.\ Rev.\ D-RC  {\bf 73}, 111101 (2006); 
\babar\ Collaboration, B.\ Aubert {\em et al.}, Phys.\ Rev.\ D   {\bf 71}, 112003 (2005)      


  \bibitem{bib:babar_sin2b}
  \babar\ Collaboration, B.\ Aubert {\em et al.}, 
  Phys.\ Rev.\ Lett.\  {\bf 94}, 161803 (2005).


  \bibitem{bib:aps-ap}
  R.~Aleksan, T.~C.~Petersen and A.~Soffer,
  Phys.\ Rev.\ D {\bf 67}, 096002 (2003); 
  R.~Aleksan and T.~C.~Petersen,
  in Proceedings of the CKM03 Workshop, Durham, 2003,
  eConf {\bf C0304052} (2003), WG414.

\bibitem{bib:nousHepPh}
F. Polci, M.-H. Schune and A. Stocchi,  
arXiv.org:hep-ph/0605129


\bibitem{bib:isobar}    
CLEO Collaboration, S. Kopp {\it et al.}, Phys.\ Rev.\ D {\bf 63}, 092001 
(2001); 
CLEO Collaboration, H. Muramatsu {\it et al.}, Phys.\ Rev. Lett. {\bf 89}, 
251802 (2002); 
erratum-ibid: {\bf 90} 059901 (2003).%



  \bibitem{pdg}
  Particle Data Group, W.-M. Yao et al., J. Phys G 33, 1 (2006).




  \bibitem{babar-nim}
  \babar\ Collaboration, B.\ Aubert {\em et al.}, 
  Nucl. Instr. Methods {\bf A479}, 1 (2002).


\bibitem{geant4}
GEANT4 Collaboration, S. Agostinelli {\it et al.},  Nucl. Instrum. Methods, {\bf A506}, 250 (2003).






 \bibitem{thrust}
  E.~Farhi,
  Phys.\ Rev.\ Lett.\  {\bf 39}, 1587 (1977).


\bibitem{argus}
  ARGUS Collaboration, H.\ Albrecht {\it et al.}, Phys.\ Lett.\ B {\bf 
185},
  218 (1987); {\it ibid.} {\bf 241}, 278 (1990).

\bibitem{bib:babarmh} 
  \babar\ Collaboration, B.\ Aubert {\em et al.}, 
Phys.\ Rev.\ Lett.\  {\bf 95}, 171802 (2005).


  \bibitem{bib:splot}
  M.~Pivk and F.~R.~Le Diberder,
  Nucl. Instrum. and Methods {\bf A555}, 356 (2005).      


\bibitem{bib:babarshahram}
\babar\ Collaboration, B.\ Aubert {\em et al.}, 
Phys.\ Rev.\ D  {\bf 74}, 031101 (2006)




\end{thebibliography}
\end{document}